\documentclass[epj]{svjour}
\usepackage{latexsym}
\usepackage[dvips]{graphics}
\usepackage{amsmath,amssymb}

\newcommand{\omt}{\Omega}
\newcommand{\om}{\omega}
\newcommand{\Bll}{B_{\lambda \lambda'}}
\newcommand{\Cpll}{C'_{\lambda \lambda'}}
\newcommand{\Cll}{C_{\lambda \lambda'}}
\newcommand{\dll}{\delta_{\lambda \lambda'}}
\newcommand{\dmm}{\delta_{m m'}}
\newcommand{\Cllbar}{\overline{C}_{\lambda \lambda'}}
\newcommand{\mCbar}{\overline{\mathbb{C}}}
\newcommand{\mCp}{\mathbb{C}\,'}
\newcommand{\vgrad}{\vec{\nabla}}
\begin{document}
\title{Stability and instability of a hot and dilute nuclear droplet}
\subtitle{II. Dissipative isoscalar modes}
\author{W.~N\"orenberg\inst{1,2} 
\and G.~Papp\inst{1,3}
\and P.~Rozmej\inst{1,4,}\thanks{\emph{e-mail addresses:} 
w.nrnbrg@gsi.de, pg@ludens.elte.hu, p.rozmej@im.pz.zgora.pl%
} }                    
%
\institute{Gesellschaft f\"ur Schwerionenforschung, 
D-64291 Darmstadt, Germany \and 
Institut f\"ur Kernphysik, Technische Universit\"at  Darmstadt,
D-64289 Darmstadt, Germany \and
HAS Research Group for Theoretical Physics, E\"{o}tv\"{o}s University,
H-1117 Budapest, Hungary
\and Institute of Physics, University of Zielona G\'ora, 
 Pl-65246 Zielona G\'ora, Poland}
\date{Received: \today }
%
\abstract{ Results on dissipative isoscalar modes of a hot and dilute nuclear droplet are presented. As compared to the adiabatic limit (part I), realistic dissipation yields a substantial reduction of the growth rates for all unstable modes, 
while the area of spinodal instability in the ($\varrho$,$T$)-plane remains unchanged. The qualitative features of multifragmentation through spinodal decomposition as obtained in the adiabatic limit are
not significantly affected by dissipation. 
} 
\maketitle
\section{Introduction and summary}\label{intro}

In the preceding part I \cite{inst} we have introduced a collective model which allows to study the eigenmodes of a nuclear droplet as function of its density $\varrho$ and temperature~$T$. We refer to this paper for a thorough
discussion of the model and related publications. The description is based on the diabatic approach to dissipative collective motion and -- in the local density approximation -- yields equations of motion for small amplitudes, where the mass and stiffness tensors are obtained analytically. The model is suited to explore systematically characteristic properties of hot nuclear droplets as function of their densities. While this is interesting by itself and particularly so in the region of the liquid-gas phase transition, such studies also yield basic information on multifragmentation by spinodal decomposition, and hence may give insights which are complementary to dynamical simulations. 

In \cite{inst} we have studied isoscalar modes in the adiabatic limit, which is defined by a vanishing relaxation time ($\tau=0$, instantaneous intrinsic equilibration). However, the adiabatic (or thermodynamic) limit is quite unrealistic for nuclear systems, because reasonable values for the relaxation time are of the same order as the characteristic times of collective motion in the region of the liquid-gas phase transition. Therefore, we extend our study here to arbitrary values 
of the relaxation time, and thereby include dissipation.

By varying $\tau$ between infinity and zero our diabatic approach allows us to treat the continuous transition between the two elastic limits, {\em i.e.} the diabatic limit($\tau \to \infty$), where the dynamical distortions of the local Fermi sphere are not destroyed by two-body collisions, and the adiabatic limit ($\tau = 0$), where the Fermi sphere is assumed to be restored instantaneously.

The method of solving the secular equation for the eigenmodes, as well as the determination of the relaxation time $\tau$ and the evaluation of the diabatic part of the stiffness coefficients (not included in \cite{inst}) are presented in section \ref{s2}. When expressed in terms of the temperature, the relaxation time of a Fermi-liquid is practically independent of the density $\varrho$, yielding rates $\hbar/\tau \approx T^2/3$MeV. In addition to the adiabatic spinodal we find also diabatic spinodals for the bulk modes of infinite nuclear matter and nuclear droplets, however not for surface modes.

Since the secular equation for the eigenmode energies is of third order, there exist three roots for $0<\tau <\infty$ instead of two for the adiabatic and diabatic limits. As functions of $\tau$ the eigenvalues move around in the
complex $\omega$-plane on generic trajectories which depend on the positions in the $(\varrho,T)$-plane (cf.~section 3).

For realistic values of the relaxation time $\tau$ the stable and unstable regions in the $(\varrho,T)$-plane are examined in section \ref{s4} and discussed in relation to the adiabatic (thermodynamic) limit of \cite{inst}. The main effects of 
dissipation are summerized as follows.
\begin{itemize}
\item The spinodal lines for bulk and surface modes are not changed by dissipation.
\item However, the growth rates of unstable bulk and surface modes are quite sensitive to dissipation and, due to the sensitivity of the relaxation time $\tau$ on temperature $T$, are reduced from their adiabatic values by factors 1/2 to 1/4 in the region of interest.
\item The qualitative differences between soft and stiff equations of state (EOS), with larger spinodal regions and larger growth rates for the stiff EOS, survive, when dissipation is included. 
\item Dissipation leads to some increase in the splitting of growth rates for modes with different multipolarities $l$ and number $n$ of radial nodes. With decreasing density the quadrupole mode $l,n=2,0$ becomes unstable first, followed in sequence by the higher multipoles $l=3,4,5$ and nodal number $n=1,2$.
\end{itemize}
The results show that the qualitative features of multifragmentation by spinodal decomposition survive when dissipation is included (section 5). 

In order to keep the presentation concise we have to refer frequently to sections, figures and equations of part I \cite{inst} and denote these references by a prefix I.

\section{Determination of dissipative eigenmodes}\label{s2}

As discussed in \cite{inst}, the eigenvalue equation for the collective modes in harmonic approximation reads (eq.~(I.19))
\begin{equation}\label{eq1}
-\Bll \,\om^2 + \Cpll\frac{\om}{\om+i/\tau} + \Cllbar  =0
\end{equation}
with the mass tensor $\mathbb{B}\equiv \{\Bll\}$, the adiabatic stiffness tensor $\overline{\mathbb{C}}\equiv \{\Cllbar\}$ and  
\begin{equation}\label{eq1a}
\mCp \equiv \{\Cpll\}= \mathbb{C} -\overline{\mathbb{C}}
\end{equation}
the difference between the diabatic ($\mathbb{C}$) and adiabatic stiffness tensors. The relaxation time $\tau$ characterizes the decay time of deformations of the local Fermi sphere.

By substituting $\om=i\omt$ one can rewrite (\ref{eq1}) as 
\begin{equation}\label{eq2}
\omt^3+\frac{1}{\tau}\omt^2 +\mathbb{B}^{-1}(\overline{\mathbb{C}}+\mCp)\omt 
+\frac{1}{\tau}\mathbb{B}^{-1}\overline{\mathbb{C}} =0 \;,
\end{equation}
where all coefficients are real numbers. This problem is equivalent (see \cite{wilk} for details) to the diagonalization of the nonsymmetric real matrix
\begin{equation}\label{matr}
\left( \begin{array}{ccc} \mathbb{O} & \mathbb{I} & \mathbb{O}\\
 \mathbb{O} & \mathbb{O} & \mathbb{I} \\ 
  ~\mathbb{B}^{-1}\overline{\mathbb{C}}/\tau~ 
& ~\mathbb{B}^{-1}(\overline{\mathbb{C}}+\mCp)~ & 
~\mathbb{I}/\tau~
 \end{array} \right) \;.
\end{equation}
All results presented in section 4 are based on this diagonalization procedure. The mass tensor $\mathbb{B}$ and the adiabatic stiffness tensor $\mCbar$ are given in \cite{inst}. In the following subsections we present expressions for the relaxation time $\tau$ and the difference $\mCp$ of the diabatic stiffness tensor with respect to the adiabatic one for 
compressional and surface modes.

\subsection{The relaxation time $\tau$}

According to \cite{bertsch} the relaxation time $\tau$ in the nuclear Fermi gas with equal numbers of neutrons and protons is given by
\begin{equation}\label{tau}
 \tau= \frac{\varepsilon_F(\varrho)}
{3.25\,\sigma\,v_F(\varrho)\,\varrho\,\varepsilon^*} \;,
\end{equation}
where $\varepsilon_F,\sigma,v_F,\varrho$ and $\varepsilon^*$ denote the Fermi energy, nucleon-nucleon cross-section without Pauli blocking, Fermi velocity, nucleon density and the excitation energy per particle, respectively. The factor 3.25 results from numerical evaluations of collision integrals. Since we are dealing with small amplitude vibrations, the energy stored in the deformation of the thermal Fermi distribution is negligible compared to the thermal energy, {\em i.e.} 
\begin{equation}\label{estar}
\varepsilon^* \approx \varepsilon_{th} = \frac{\pi^2\,T^2}
{4\varepsilon_F(\varrho)} = 
\alpha_0\,T^2 \left(\frac{\varrho_{eq}}{\varrho}\right)^{2/3}
\frac{m^*(\varrho)}{m^*(\varrho_{eq})} \;,
\end{equation}
where $\varepsilon_F(\varrho)\propto\varrho^{2/3}m^*(\varrho_{eq})/m^*(\varrho)$ and the reduced level-density 
parameter $\alpha_0 = a_0/A$ for stable nuclei with equilibrium density $\varrho_{eq}$ has been introduced. Replacing $\varepsilon^*$ in (\ref{tau}) by the expression (\ref{estar}) we find
\begin{equation}\label{tauh}
 \frac{\tau}{\hbar} = \frac{\eta}{\alpha_0\,T^2}
                \frac{m^*(\varrho)}{m^*(\varrho_{eq})}
\end{equation}
with the $\varrho$-independent and dimensionless constant 
\begin{equation}\label{eta}
 \eta = \frac{(3\pi^2/2)^{1/3}}{6.5}\,\frac{1}{\sigma\,\varrho_{eq}^{2/3}}
 = 0.36 \;,
\end{equation}
where $\sigma=40$ mb (4 fm$^2$) and $\varrho_{eq}=0.14$ fm$^{-3}$ has been used. Further realistic values for $\eta$ have been reported in \cite{kolom} and range from 0.25 to 0.52 for $\alpha_0= 0.1$ MeV$^{-1}$. Note that the range is roughly [0.15$\ldots$0.30] if the Fermi-gas value 0.067 for $\alpha_0$ is applied \cite{inst}. Through $m^*$ the relaxation time depends only weakly on the density $\varrho$. In the numerical calculations of stability and instability (section 4) $\eta=0.25$ is used, hence we have for the relaxation rate typically $\hbar/\tau \approx T^2/3$MeV in the spinodal region, where $m^*(\varrho)/m^*(\varrho_{eq})\approx 1.2$.

\subsection{Stiffness tensor $\mCp$}

According to (\ref{eq1a}) $\mCp$ denotes the difference between the diabatic and adiabatic stiffness tensors. Since their interaction parts are identical, $\mCp$ reduces to the difference 
\begin{equation}\label{cprim}
\mCp = \mathbb{C}^{(\cal T)} - \mCbar^{(\cal T)}
\end{equation}
of the diabatic and adiabatic contributions from the intrinsic kinetic energy. In evaluating this difference in the local density approximation we start from the corresponding expression for the diabatic intrinsic kinetic energy
\begin{equation}\label{kine}
E^{intr}_{kin} \!=
\! \int\! d^3k \,f(\vec{k})\!\int\! d^3r' 
\frac{\hbar^2 \vgrad'\widetilde{\Phi}^*_\vec{k}(\vec{r'}) \vgrad'\widetilde{\Phi}_\vec{k}(\vec{r'})}
{2m^*(\widetilde{\varrho}(\vec{r'}))} 
\end{equation}
for protons and neutrons, separately. Here $\vec{r'}=\vec{r}+\vec{s}(\vec{r})$ is the position shifted by the displacement field $\vec{s}(\vec{r})$ and $\widetilde{\Phi}_\vec{k}(\vec{r'})$, $\widetilde{\varrho}(\vec{r'})$ denote, respectively,  the distorted normalized plane-wave functions and density at the point $\vec{r'}$. The momentum distribution $f(\vec{k})$ remains fixed in the diabatic limit and is normalized to the density $\varrho$ of the unperturbed homogenous droplet. With (I.A.3) and (I.A.8) expression (\ref{kine}) in the local-density approximation reduces to 
\begin{figure}[t]
\begin{center}
 \resizebox{0.46\textwidth}{!}{\includegraphics{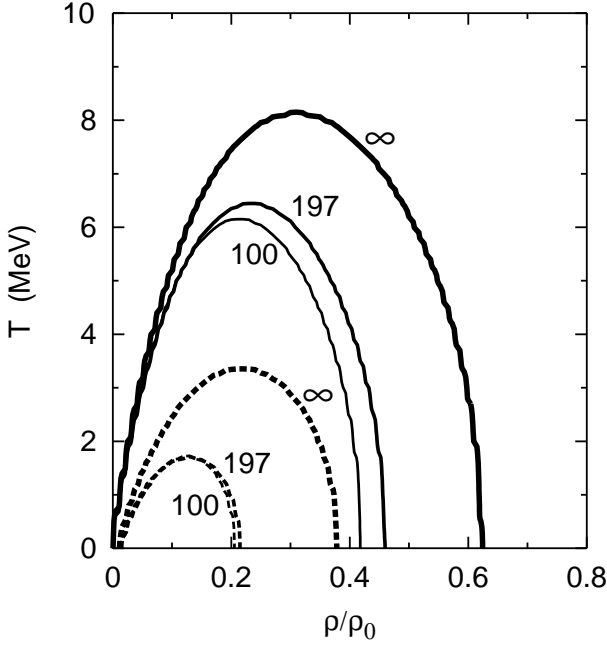}} 
\end{center}
\caption{Adiabatic (solid lines) and diabatic (dashed lines) spinodals in the $(\varrho,T)$-plane for infinite nuclear matter (indicated by $\infty$), gold-like ($A$=197, $Z$=79, $N$=118) and tin-like ($A$=100, $Z$=$N$=50) nuclear droplets  calculated for a soft EOS (SkM$^*$). Here, $\varrho_0=0.16$ fm$^{-3}$ is the normal nuclear matter density. Note that
the equilibrium density of the droplets is around $0.85\varrho_0$.}
\vspace*{-2mm}
\label{inf}       
\end{figure}
\begin{equation}\label{kine1}
E^{intr}_{kin} = \varrho \int_{r<R} d^3r 
\frac{\hbar^2 \langle \widetilde{\vec{k}}^2\rangle_{diab}} 
{2m^*(\widetilde{\varrho})} \;,
\end{equation}
where $\widetilde{\varrho}$ is given by (I.A.12). The distorted momenta $\widetilde{\vec{k}}$ are determined according to (I.A.3) by 
\begin{equation}\label{aint2}
 \widetilde{k_j} = \sum_i\, k_i\frac{\partial r_i}{\partial {r'}_j}
\end{equation}
with ${r'}_j=r_j+s_j(\vec{r})$, and hence
\begin{equation}\label{aint3}
 \langle \widetilde{\vec{k}}^2 \rangle = \int \,\mathrm{d}^3 k\, 
 f(k)\,\widetilde{\vec{k}}^2 = \frac{1}{3}\langle \vec{k}^2 \rangle
 \sum_{i,j} \left( \frac{\partial r_i}{\partial {r'}_j} \right)^2 \;,
\end{equation}
where the isotropy of $f(k)$ has been used. Inserting the expansion
\begin{equation}\label{aint4}
 \frac{\partial r_i}{\partial {r'}_j} = \delta_{ij} - \sum_{k}\,
 \frac{\partial s_i}{\partial r_k} \left( \delta_{kj} - 
 \frac{\partial s_k}{\partial r_j}\right) + {\cal O}(s^3)
\end{equation}
up to second order in $\vec{s}$, we obtain with (I.41)
\begin{equation}\label{aint5}
 \langle \widetilde{\vec{k}}^2 \rangle_{diab} = \langle \vec{k}^2 \rangle
 \{ 1 -\frac{2}{3}\mathrm{div}\,\vec{s} + \sum_{i,j}
 (\partial_i \partial_j\, w)^2 \}
\end{equation}
in the diabatic limit.

With the adiabatic expression for $\langle \widetilde{\vec{k}}^2\rangle$ from \cite{inst}
\begin{equation}\label{kadiab}
 \langle \widetilde{\vec{k}}^2\rangle_{adiab} = 
 \langle k^2\rangle\{1- \frac{2}{3}\mathrm{div}\,\vec{s} 
 +\frac{2}{9}(\mathrm{div}\,\vec{s})^2 + \frac{1}{3} \sum_{i,j}
 (\partial_i \partial_j\, w)^2 \}
\end{equation}
(note the misprint in (I.C.2)) we finally can evaluate $\mCp$ according to (\ref{cprim}) as
\begin{eqnarray}\label{cpri}
\Cpll & = & \varrho \int d^3r \,\frac{\hbar^2\langle k^2\rangle}{2m^*(\varrho)}
\\ & & \times
\frac{\partial^2}{\partial q_\lambda\partial q_{\lambda'}}
 \left( \frac{2}{3} \sum_{i,j}
 (\partial_i \partial_j\, w)^2 - \frac{2}{9}(\mathrm{div}\,\vec{s})^2
\right) \;, \nonumber
\end{eqnarray}
where the density dependence of $m^*$ does not contribute, because the bracket is already of second order in $q_\lambda$.

\subsubsection{Compression modes}

The displacement fields for the compressional modes have been defined by (I.34,I.41) and discussed in detail in \cite{inst},
\begin{equation}\label{gf8}
 \vec{s}(\vec{r},t) = \vgrad \, w(\vec{r},t) = 
 \sum_{\lambda}\,q_{\lambda}(t)\,\vgrad\chi_{\lambda}(\vec{r}) \; .
\end{equation}
with the normalized functions
\begin{equation}\label{gf1}
\chi_{\lambda \equiv \{nlm\}}(\vec{r}) = {\cal N}_{nl} \, j_l(\kappa_{nl}r) \, {\cal Y}_{l}^{m}(\Omega) \; , 
\end{equation}
where $j_l(\kappa_{nl}r)$ denote the spherical Bessel functions with $j_l(\kappa_{nl}R)=0$, and ${\cal Y}_{l}^{m}(\Omega)$ 
the real spherical harmonics. 

By applying the relations (I.C.4) to (I.C.7), noting a wrong sign in (I.C.6), we obtain from (\ref{cpri}) the final result 
\begin{equation}\label{cpri1}
\Cpll \! = \! \sum_{i=\mbox{\scriptsize n,p}} 2\epsilon^{(i)}_{\cal T}(\varrho,T)\frac{2}{3} 
\left\{ \frac{2}{3} \kappa^4_{nl}\dll - 
\frac{4\kappa_{nl}\kappa_{n'l'}}{R^2} \delta_{ll'} \dmm \right\} ,
\end{equation}
where the contributions for protons ($i=$p) and neutrons ($i=$n) are summed. 

When comparing the diagonal elements $C'_{\lambda\lambda}$ with $\overline{C}_{\lambda\lambda}^{({\cal T})}$, {\em i.e.}  with the contribution from the intrinsic kinetic energy to the adiabatic stiffness coefficient (cf.\ eq.~(I.46)), one realizes that  $C'_{\lambda\lambda}$ is of almost the same magnitude, also always positive, because $(\kappa_{nl}R)^2\gg 1$, and hence a major contribution to the diabatic stiffness coefficient $C_{\lambda\lambda}=C'_{\lambda\lambda}+ \overline{C}_{\lambda\lambda}\,,$ cf.\ figs.~I.3 and I.4. As a consequence the diabatic spinodal in the $(\varrho,T)$-plane, which essentially is determined by $\Cll (\varrho,T)=0$ with $l=2,n=0$, lies well inside the adiabatic spinodals as shown in fig.~\ref{inf} for infinite nuclear matter and two nuclear droplets with A=197 and 100. Like for the adiabatic spinodals the large finite size effect is due to the finiteness of the minimum $\kappa_{nl}$-value for the droplets. In the diabatic limit this almost leads to disappearance of the spinodal region. Furthermore, the diabatic spinodals for the gold-like and tin-like droplets are almost identical like in the adiabatic limit. 

\subsubsection{Pure surface modes}

In accordance with \cite{N&S} and (I.59) we define the displacement field for pure surface modes by
\begin{equation}\label{dispf}
\vec{s}(\vec{r},t) = 
 \sum_{l\ge 2,m}\frac{Q_{lm}}{l\,R^{l-2}}\vgrad (r^l{\cal Y}_l^m) \;,
\end{equation}
which is consistent with incompressibility of the flow up to first order in the collective variables $Q_{lm}$. This is sufficient for the evaluation of (\ref{cpri}) up to second order in $Q_{lm}$, and hence with $\vgrad \cdot \vec{s}=0$ we have
\begin{equation}\label{cpris}
\Cpll = \sum_{i=n,p} \, 2\epsilon^{(i)}_{\cal T}(\varrho,T)
 \frac{2}{3} \int d^3r
 \sum_{i,j} (\partial_i\partial_j w)^2 \;.
\end{equation}

Taking the additional factor $\sqrt{(2l+1)/4\pi}$ in the definition of our collective variables into account, we find from \cite{N&S}
\begin{equation}\label{ddw}
\int d^3r \sum_{i,j} (\partial_i\partial_j w)^2 = 
 \frac{(l-1)(2l+1)}{2l}R^3 
\end{equation}
 and
\begin{equation}\label{e21}
\Cpll = \sum_{i=n,p} \, 2\epsilon^{(i)}_{\cal T}(\varrho,T) 
 \frac{(l-1)(2l+1)}{3l}R^3 \;.
\end{equation}
For completeness we also give the adiabatic stiffness tensor for the surface modes 
\begin{equation}\label{adsti}
 \mCbar = \mathbb{C}^{(S)} + \mathbb{C}^{(C)}
\end{equation}
with
\begin{equation}\label{adstiS}
 \mathbb{C}^{(S)} = \mathbb{C}^{(S)}_{BM} 
 \frac{\varepsilon_{S}(\varrho,T)}{\varepsilon_{S}(\varrho_{eq},0)}
 \left(\frac{\varrho_{eq}}{\varrho}\right)^{2/3} \;, 
\end{equation}
\begin{equation}\label{adstiC}
\mathbb{C}^{(C)} = \mathbb{C}^{(C)}_{BM} 
 \left(\frac{\varrho}{\varrho_{eq}}\right)^{1/3} \;, 
\end{equation}
where $\mathbb{C}^{(S,C)}_{BM}$ denote the expressions (I.63) and (I.64) from \cite{inst} and $\varrho_{eq}$ the density of the cold stable droplet. The $\varrho$- and $T$-dependence have not been reported in \cite{inst}.
No diabatic spinodal exists for the surface modes because $C'\gg |\overline{C}|$.

\section{Generic features of dissipative modes}\label{s3}

For simplicity we consider a single mode such that the matrix equation (1) reduces to 
\begin{equation}\label{sme}
 G^{-1}(\omega) = - B\omega^2 + C'\frac{\omega}{\omega+i/\tau} +
 \overline{C} =0 \;,
\end{equation}
where $G(\omega)$ denotes the Green function. Since the coupling of different modes does not alter the qualitative dependence of the three roots (poles of the Green function) on the parameters $B,C',\overline{C}$ and $\tau$, we are able to reveal generic features of the dissipative modes from (\ref{sme}).

\subsection{Motion of poles as function of $\tau$}   

\begin{figure}[t]
\begin{center}
\vspace*{0mm}
 \resizebox{0.48\textwidth}{!}{\includegraphics{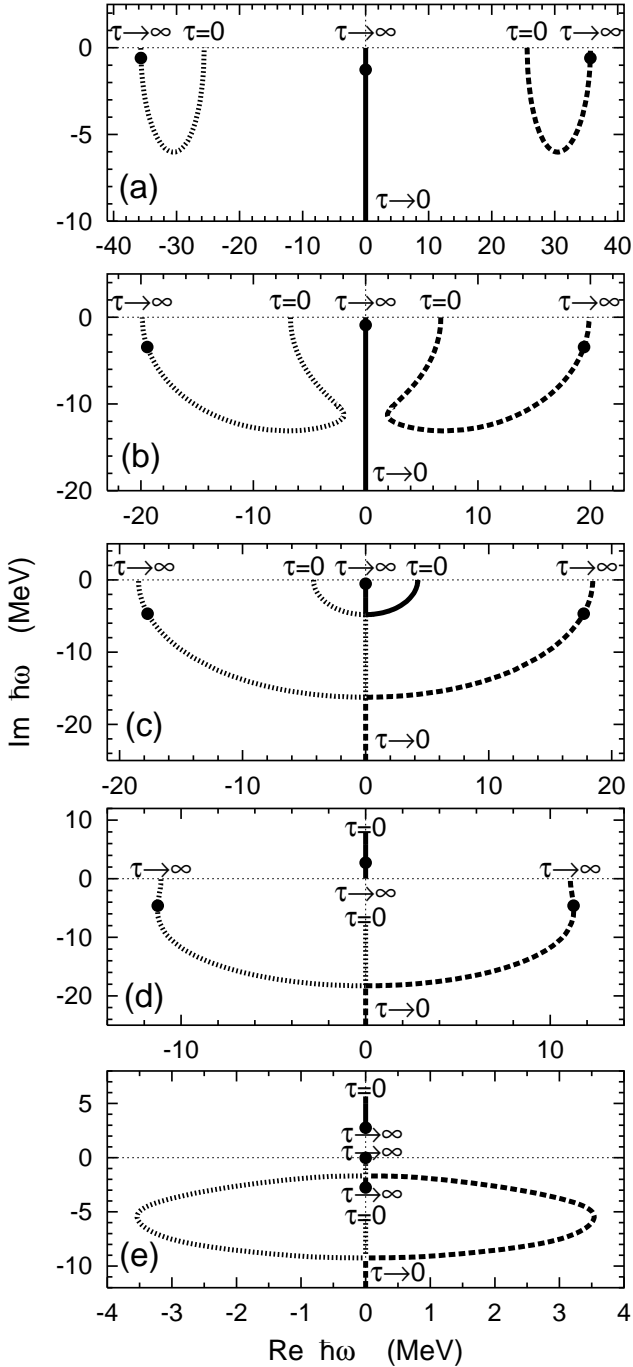}} 
\end{center}
\vspace*{-2mm} 
\caption{Generic behavior of the eigenvalues of eq.~(\ref{sme}) in the complex $\omega$-plane as functions of the relaxation time $\tau$, presented in (a) to (e) for different regions of stability and instability (different values of $\overline{C}$ and $\overline{C}+C'$, see text). Limits for $\tau\rightarrow 0$ and $\tau\rightarrow\infty$ are indicated. The bullets on the trajectories correspond to realistic $\tau$-values obtained for $\eta=0.25$.}
\vspace*{-4mm}
\label{modtau}       
\end{figure}

For different values of the mass and stiffness parameters we illustrate these properties by the motion of the roots (poles) in the complex $\omega$-plane as function of $\tau$, which is freely varied from the diabatic limit $(\tau\rightarrow\infty)$ to the adiabatic limit $(\tau\rightarrow 0)$, thereby creating trajectories for the three poles in the complex $\omega$-plane. For the presentation in fig.~\ref{modtau} we have chosen the compressional mode $l,n = 2,0$ of a gold-like droplet for the soft EOS (SkM$^*$) at five points in the $(\varrho,T)$-plane (cf. fig.~\ref{inf}), ranging from the region of stable nuclei via the vicinity of the adiabatic spinodal around $T=3$ MeV into the regions of adiabatic and diabatic spinodal instabilities. 

\begin{itemize}
\item 
The top part (a) of fig.~\ref{modtau} corresponds to a gold nucleus at small temperatures $(C',\overline{C}>0)$. For   $\tau\rightarrow\infty$ there exist essentially two poles of diabatic vibration. In this limit the third pole corresponds to an amplitude constant in time added to the time-dependent vibration. For increasing dissipation, {\em i.e.} decreasing $\tau$-values, the diabatic poles become complex, moving into the lower half of the complex $\omega$-plane, and hence correspond to damped oscillations. The third pole (sometimes called thermal pole) moves also down on the imaginary axis. For $\tau\rightarrow 0$ the two vibrational poles approach their adiabatic positions on the real axis while the thermal pole quickly escapes to $-i\infty$. 
\item
When approaching the adiabatic spinodal $(C'>0,\overline{C}\gtrsim 0)$ around $T=3$ MeV, we observe an interesting deformation of the pole trajectories as illustrated in part (b) and (c). The vibrational pole trajectories approach each other and in  (b) almost touch.  Only somewhat closer to the spinodal line the trajectories are no longer separated, and hence the poles loose their original physical meaning. Starting from the diabatic limit ($\tau\rightarrow\infty$) the two vibrational poles meet on the negative imaginary axis and then move up and down (towards $-i\infty$) while the thermal pole moves down slowly on the imaginary axis. The upward moving ``diabatic" pole collides with the downward moving ``thermal" pole and finally approach the adiabatic pole positions as $\tau\rightarrow 0$.
\item
The trajectories for a position somewhat inside the adiabatic spinodal $(\overline{C}\lesssim 0)$, but outside the diabatic spinodal $(C=C'+\overline{C}>0$) are shown in part (d). Now for all finite values of $\tau$ there is a pole on the positive imaginary axis reflecting the spinodal instability encountered in the spinodal region. The two diabatic poles move down with decreasing $\tau$ and, after colliding on the negative imaginary axis, they approach the adiabatic vibrational position in the lower half plane and $-i\infty$, respectively.
\item 
For a position inside the diabatic spinodal (part (e)) both $\overline{C}$ and $C=C'+\overline{C}$ are negative, such that both pairs of adiabatic and diabatic vibrational poles lie on the imaginary axis. With decreasing $\tau$-values the instability pole moves again upwards on the positive imaginary axis, however starting now from the finite value in the diabatic limit  $\tau\rightarrow\infty$ and reaching the adiabatic value for $\tau\rightarrow 0$. For decreasing $\tau$ the thermal pole collides with the upward moving lower diabatic pole and after a detour in the lower complex half plane both poles collide again on the imaginary axis and finally move to the adiabatic position in the lower half plane and to $-i\infty$, respectively.
\end{itemize}

\begin{figure}[b]
\begin{center}
 \resizebox{0.48\textwidth}{!}{\rotatebox{-90}
              {\includegraphics{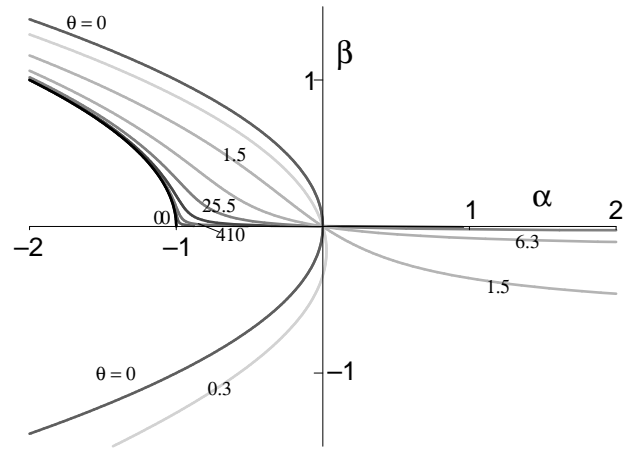}}} 
\end{center}
\caption{Universal curves for the evolution of the instability poles $\beta = \gamma\sqrt{B/C'}$ as functions of $\alpha = \overline{C}/C'$ for different relaxation times $\tau$ between the adiabatic ($\tau = 0$) and diabatic ($\tau \rightarrow \infty$) limits: $\theta = \tau \sqrt{C'/B}$ = 0, 0.3, 1.5, 6.3, 25.5, 102, 410, 1638, $\infty$.}
\vspace*{-2mm}
\label{instpole}       
\end{figure}

\subsection{The pole of instability}

We look for the pole on the imaginary axis $\omega = i\gamma $ for finite values of $\tau$. According to (\ref{sme}) we have 

\begin{equation}\label{instpole1}
   B\gamma^2 + C'\frac{\gamma}{\gamma+1/\tau} + \overline{C} =0 \;, 
\end{equation}
which for the dimensionless quantities $\alpha = \overline{C}/C'$ , $\beta = \gamma \sqrt{B/C'}$ and $\theta = \tau \sqrt{C'/B}$ reads 
\begin{equation}\label{instpole2}
   \alpha = - \beta^2 - \frac{\beta}{\beta+1/\theta}  \;. 
\end{equation}
This universal function is illustrated in fig.~\ref{instpole} for $\beta > - 1/\theta$, {\em i.e.} above the singularity  $\beta = - 1/\theta$. All curves enter the region of instability $\beta > 0$ through the origin, where the adiabatic stiffness coefficient $\overline{C}$, and hence $\alpha$  becomes negative. The family of curves map the whole region between the adiabatic limit $\tau = 0$ and the diabatic limit $\tau \rightarrow \infty$ ($\theta \rightarrow \infty$). In the unstable region between $-1<\alpha <0$ the growth rates are quite sensitive to the relaxation time and vanish in the diabatic limit. 

\begin{figure}[t]
\begin{center}
 \resizebox{0.38\textwidth}{!}{ \includegraphics{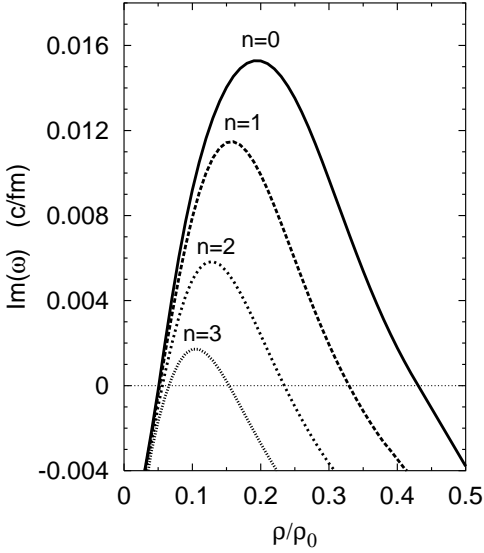}  } 
\end{center}
\caption{Growth rates (Im($\omega) > 0$) of the unstable compressional quadrupole modes ($l=2$) as functions of density at $T=3$ MeV for a gold-like droplet and the soft EOS. Since the coupling between the radial modes is small, the eigenmodes are still well characterized by the nodal numbers $n=0,1,2,3$.}
\label{T3i}      
\end{figure}

\section{Effects of dissipation on stability and instability}\label{s4}

Eigenvalues are obtained by numerical diagonalization of the matrix (\ref{matr}) for given values of density and temperature. As pointed out in \cite{inst}, modes with different angular momenta $l$ are decoupled. Coupled are modes with the same $l$ and different $n$ ($n$ is the number of nodes in the spherical Bessel function determining the radial density profile). As the energy of modes grows substantially with $n$, while the couplings between the modes is relatively small, it is sufficient to take into account only a few modes with the smallest $n$-values. We have studied all modes with $l=2,3,4,5$ and included in the diagonalization the radial modes with $n=0,1,2,3$ for each $l$-value. As mentioned in section 2.1 we use $\eta = 0.25$ for calculating the relaxation time $\tau$. 

\subsection{Compression (bulk) modes}

\begin{figure}[t]
\begin{center}
 \resizebox{0.38\textwidth}{!}{ \includegraphics{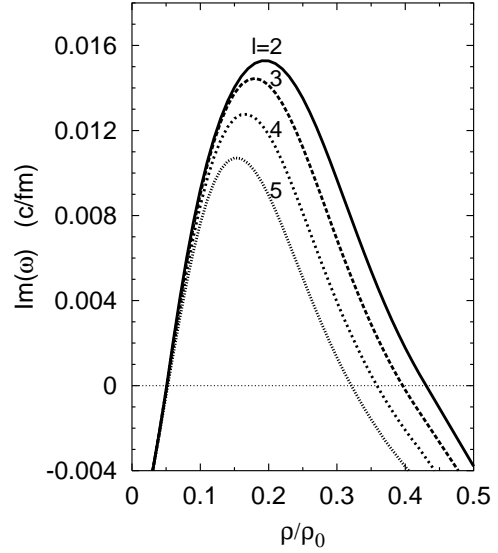}  } 
\end{center} 
\caption{Growth rates (Im($\omega) > 0$) of the unstable compression modes with multipolarities 
$l=2,3,4,5$ as functions of density at  {$T=3$~MeV} for a
gold-like droplet and the soft EOS. Note that the curve $l=2$ is identical with the curve $n=0$ of fig.~\ref{T3i}.}
\label{lmI}       
\end{figure}

The results on the compression modes are presented in figs.~\ref{T3i} to \ref{l2-4soft}. 

\subsubsection{Dependence on $n$ and $l$}

Like  in the adiabatic limit (cf.~figs.\ I.5 to I.7) more and more modes with $n>0$ for a given $l$-value become unstable  (Im($\omega) > 0$) with decreasing density as illustrated in fig.~\ref{T3i} for $l=2$. However, dissipation reduces the magnitude of the growth rates by factors  1/2 to 1/4 and, in contradistinction to the adiabatic limit, no crossings of different modes as function of $\varrho$ is observed. Furthermore, the curves are not that well degenerate at small densities as in the diabatic limit, such that the modes with $n=0,1$ dominate. This behavior is typical for all multipole modes as well as for the soft and stiff EOS.

Figure \ref{lmI} (the analog to fig.~I.6) shows the largest growth rates (Im($\omega) > 0$) for different multipoles ($l=2,3,4,5$). With decreasing density the growth rates for different $l$-values come close to each other. However, again the degeneracy is not as perfect as in the adiabatic limit. Still, different multipolarities will strongly compete in the decay of the droplet.

\subsubsection{Dependence on $\varrho$ and $T$}
\begin{figure}[t]
\begin{center}
 \resizebox{0.43\textwidth}{!}{\includegraphics{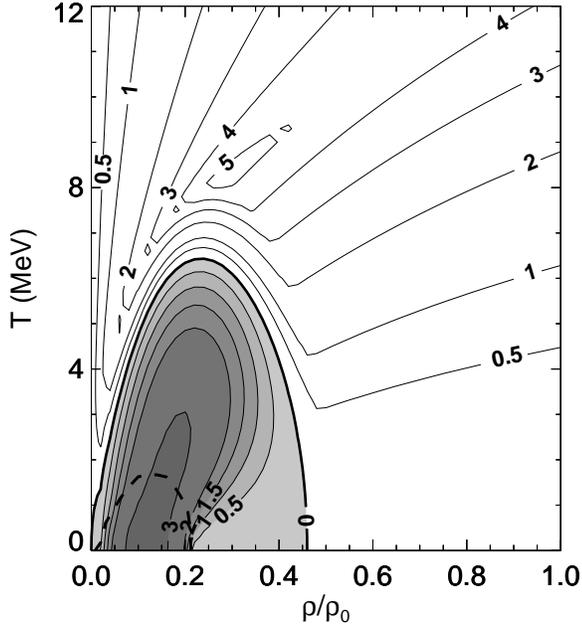}} 
\end{center}
\caption{Contour plot of largest growth rates in the unstable (shaded) region and smallest vibrational damping rates in the stable region for dissipative compression modes of a gold-like droplet described by the soft EOS.
The numbers on the lines denote the rates Im($\hbar\omega$) in MeV. All modes with $l=2,3,4$ and $n=0,1,2,3$ are taken into account. The dashed line indicates the diabatic spinodal.}
\label{l2-4soft}       
\end{figure}
\begin{figure}[t]
\begin{center}
 \resizebox{0.43\textwidth}{!}{\includegraphics{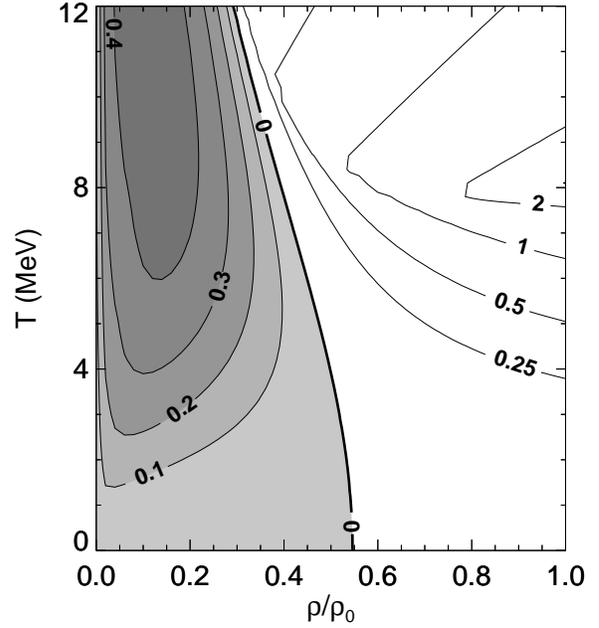}} 
\end{center}
\caption{Contour plot of largest growth rates in the unstable (shaded) region and smallest damping rates in the stable region for dissipative surface modes of a gold-like droplet described by the soft EOS. All modes with $l=2,3,4$ are taken into account.}
\label{l2-4surf}       
\end{figure}
In contradistinction to \cite{inst} we present in the regions of stability instead of the roots $\omega$ only the damping rates $\gamma$ = Im($\omega) < 0$ of the vibrations. These quantities are relevant for the restoration of the equilibrium shapes. In the spinodal regions (the regions of instability) the positive imaginary roots determine the growth rates $\gamma$ = Im($\omega) > 0$ of the unstable modes. The following discussion of stability and instability is based on these quantities.

Figure \ref{l2-4soft} (the analog to fig.~I.7) shows the largest growth rates in the unstable (shaded) region and the smallest damping rates in the stable region for a gold-like droplet described by the soft EOS. The adiabatic spinodal line is not affected by dissipation, because it is essentially determined by $\omega_{adiab}$ for $l,n=2,0$, the coupling from finite $\tau$-values between different $n$ being small in the relevant regions $\varrho \approx 0.25\varrho_{eq}$, $T \approx 4$~MeV. There the growth rates are reduced by factors 1/2 to 1/4 as mentioned  before. Furthermore, a pronounced effect from the diabatic instability is seen for small $T$ (large $\tau$-values), consistent with the behavior of the growth rates around $\alpha = -1$ in fig.~\ref{instpole}. Since $\tau \propto 1/T^2$ the growth rates and damping rates approach zero on the real axis to the right of the diabatic spinodal.
\begin{figure}[t]
\begin{center}
 \resizebox{0.43\textwidth}{!}{\includegraphics{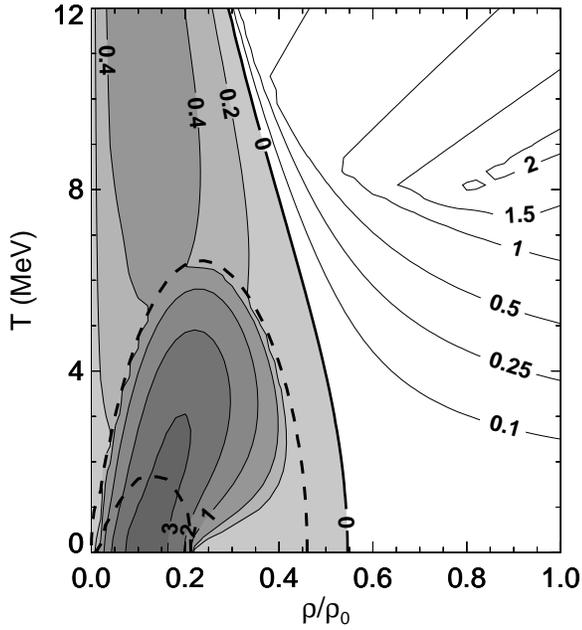}} 
\end{center}
\caption{Combined bulk and surface instabilities (soft EOS) for a gold-like droplet. The dashed lines indicate the adiabatic and diabatic spinodals.}
\label{l2-4busu}       
\end{figure}

\begin{figure}[t]
\begin{center}
 \resizebox{0.43\textwidth}{!}{\includegraphics{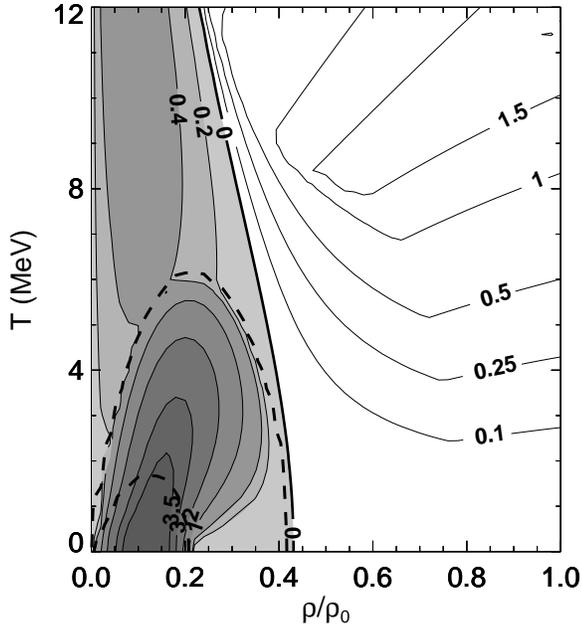}} 
\end{center}
\caption{The same as in fig.\ \ref{l2-4busu}, but for a tin-like droplet.}
\label{l2-4tin}       
\end{figure}

\subsection{Surface modes}

Figure \ref{l2-4surf} (the analog to fig.~I.8) summarizes the results for the surface modes. As for the bulk modes, dissipation reduces the growth rates of instabilities by factors 1/2 to 1/4 in the region of interest ($T \approx 5$ MeV).  The spinodal line is again not affected by dissipation. The shift in fig.~\ref{l2-4surf} to smaller densities as compared to fig.~I.8 is not an effect from dissipation but due to the inclusion of the correct density dependence also in the Coulomb part (25) of the stiffness coefficient, which was neglected in \cite{inst}.

\subsection{Combined plot of bulk and surface modes}

A combined plot of compression (bulk) and surface modes is presented in fig.~\ref{l2-4busu} (the analog to fig.~I.9). Like in the adiabatic limit the compressional instabilities dominate at small densities and temperatures, while for large
temperatures and densities outside the bulk spinodal only surface modes are unstable.

\subsubsection{Dependence on size}

Figure \ref{l2-4tin} (the analog to fig.~I.10) shows the results in the $(\varrho,T)$-plane for the tin-like droplet. Like in the adiabatic limit the compression instabilities of a finite nuclear droplet depend only weakly on the mass.  However,  the surface instabilities are pushed to considerably smaller densities for the lighter system.

\begin{figure}[t]
\begin{center}
 \resizebox{0.43\textwidth}{!}{\includegraphics{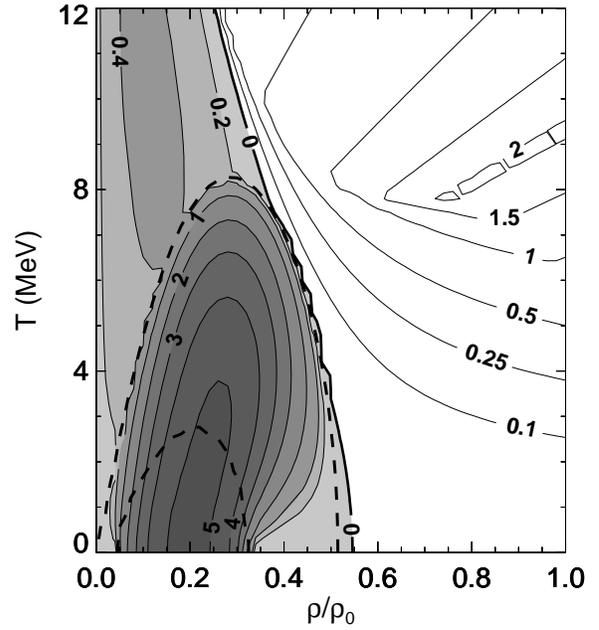}} 
\end{center}
\caption{The same as in fig.\ \ref{l2-4busu}, but for the stiff EOS.}
\label{l2-4busustif}       
\end{figure}

\begin{figure}[t]
\begin{center}
 \resizebox{0.43\textwidth}{!}{\includegraphics{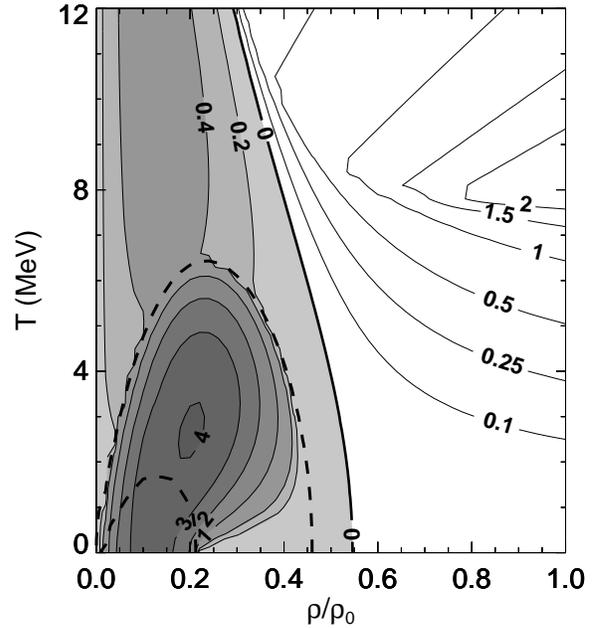}} 
\end{center}
\caption{The same as in fig.\ \ref{l2-4busu}, but for smaller relaxation times with $\eta=0.12$.}
\label{l2-4busuet015}       
\end{figure}

\subsubsection{Dependence on EOS}

Figure \ref{l2-4busustif} (the analog to fig.~I.11) displays the combined instabilities for the gold-like droplet and the stiff EOS. Dissipation does not alter the qualitative dependence on the EOS. The region and degree of compressional instability is considerably larger for the stiff EOS. Again the growth rates in the relevant region are reduced by factors  1/2 to 1/4 due to dissipation. As far as the calculations are comparable these results are consistent with those quoted in \cite{K&S}.

\subsubsection{Sensitivity to dissipation}

Figure \ref{l2-4busuet015} illustrates the sensitivity of the growth rates on the reduction of the relaxation time to half of its value. While the surface instabilities remain almost unchanged, the bulk instabilities gain roughly 30\% in the growth rates. Since a factor of 2 defines roughly the uncertainty range for the relaxation time we can consider the growth rates  presented in figs.~\ref{l2-4soft} to \ref{l2-4busustif} reliable to about $\pm$30\%.

\section{Effects on multifragmentation by spinodal decomposition}\label{s5}

In section I.5 we have shown how spinodal decomposition through adiabatic surface and bulk instabilities may explain qualitatively and to some extent even quantitatively characteristic features of multifragmentation reactions. In discussing the effects of realistic relaxation by two-body collisions we consider the following three consecutive stages of multifragmentation:
\begin{itemize}
\item[--]the initial formation of a hot nuclear droplet with density $\varrho \approx \varrho_{eq}$,
\item[--]expansion into the spinodal region and
\item[--]spinodal decomposion.
\end{itemize}

The various reactions (cf. \cite{inst}), which are studied with respect to multifragmentation, yield nuclear droplets typically with excitation energies $E^*/A \approx 6$ MeV. At these high excitations two-body collisions lead to local intrinsic equilibration (relaxation of local Fermi sphere) on short time scales of $\tau \approx 10$~fm/c, cf.~section 2.1. Thus thermal intrinsic equilibrium is attained already at the beginning of the expansion. However, this equilibration does not include the relaxation of deformations which are produced in the formation of the heated droplet. As shown in figs.~\ref{l2-4soft} and \ref{l2-4surf} bulk and surface modes relax on the time scales of $\hbar /$(2~MeV) $\approx 100$~fm/c and $\hbar /$(1~MeV) $\approx 200$~fm/c, respectively, and hence survive to a large extent the expansion process into the spinodal region (duration time $\approx$ (30\ldots50)~fm/c, cf.~\cite{papp}). Therefore, in order to study for example the bulk instabilities in the spinodal region, it is important to create the hot droplet in a gentle way without too large density distortions. Otherwise the surviving initial density fluctuations could mask the characteristics of spinodal decomposition.

The expansion of the heated droplet is fast, as mentioned before. However, no significant deviations from intrinsic equilibrium is expected. The reason is that the expansion is well approximated by the radial flow $\vec{v}(\vec{r})= a\vec{r}$, which keeps the density homogeneous within the droplet. This flow scales the single particle energies like the temperature with $\varrho^{2/3}/m^*(\varrho)$ such that the product of single-particle energy and temperature, and hence the single-particle occupation probabilities remain constant. Thus starting from an equilibrium distribution, the system stays during this isentropic expansion in equilibrium even without relaxation by two-body collisions. Consequently, within the collective model described by (\ref{eq1}), the diabatic and adiabatic stiffness coefficients are identical, and hence no damping of the expansion occurs ($C'=0$). Dynamical calculations \cite{papp} have shown that for not too large initial excitations ($E^*/A \lesssim 10$~MeV) the expansion reaches turning points at temperatures around 5~MeV which lie well inside the spinodal line.

As discussed in section 3 and 4, the spinodal regime does not deviate from the one obtained in the adiabatic limit. Although the growth rates for the spinodal instabilities are considerably reduced when realistic two-body collisions are taken into account, the decay times are still short enough such that spinodal decomposition occurs with\-in the times the droplet remains around the turning point inside the spinodal line. Furthermore, while the density fluctuations grow, the restoring force of the radial flow is reduced, and hence the expansion tends to continue such that the spinodal decomposition is likely to complete. Therefore, we expect the qualitative features of multifragmentation as discussed in section I.5 to survive for realistic dissipation.  

During the clustering process thermalization within the decaying droplet is so fast (relaxation time $\tau \approx$ 20~fm/c for $T \approx 5$~MeV)  that statistical equilibrium is attained throughout the formation of fragments. This explains why statistical models are so successful in describing various aspects of multifragmentation. This may change as soon as the expansion is not completely slowed down and large flow velocities tear the droplet dynamically apart without allowing complete equilibration \cite{reisd,shik}. 

Recently, evidence of bulk fragmentation and spinodal decomposition have been reported in mass distributions for the collisions $^{155}$Gd(36~MeV/u)+U and $^{129}$Xe(32 MeV /u)+Sn, cf.~\cite{frank,bord}. However, since mass distributions are dominated by statistical decay probabilities, the identification of spinodal decomposition turns out to be difficult. The situation may improve when the onset of spinodal decomposition is studied by looking into excitation functions.

\begin{acknowledgement}
\subsection*{Acknowledgements}
We gratefully acknowledge fruitful discussions with our experimental 
colleagues 
U.~Lynen, W.~Reisdorf, C.~Schwarz and W.~Trautmann.
\end{acknowledgement}


\begin{thebibliography}{99}

\bibitem{inst}
 W.~N\"orenberg, G.~Papp and P.~Rozmej, Eur.\ Phys.\ J.\ A \textbf{9}, 327 (2000) and {\em references therein}
 
\bibitem{wilk}
 J.H.~Wilkinson \textit{The Algebraic Eigenvalue Problem} (New York, Oxford University Press, 1965)

\bibitem{bertsch}
 G.F.~Bertsch, Z.\ Phys.\ A  \textbf{289}, 103 (1978)

\bibitem{kolom}
V.M.~Kolomietz, V.A.~Plujko and S.~Shlomo, Phys.\ Rev.\ C \textbf{52}, 2480 (1995)
 
\bibitem{N&S}
J.R.~Nix and A.J.~Sierk, Phys.\ Rev.\ C  \textbf{21}, 396 (1980)

\bibitem{B&M}
A.~Bohr and B.R.~Mottelson, \textit{Nuclear Structure} (Ben\-ja\-min, Mas\-sachusetts 1975) Vol.II, Appendix 6A
 
\bibitem{K&S}
V.M.~Kolomietz and S.~Shlomo, Phys.\ Rev.\ C \textbf{60}, 044612 (1999)
 
\bibitem{papp}
G.~Papp and W.~N\"orenberg, APH Heavy Ion Physics {\bf 1}, 241 (1995);
W.~N\"orenberg and G.~Papp,  \textit{Critical Phenomena and Collective Observables}, ed. by S.~Costa,  S.~Albergo, A.~Insola and C.~Tuve (World Scientific, Singapore 1996) p.\ 377, and {\em to be published}

\bibitem{reisd}
W.~Reisdorf {\em et.~al.} (FOPI collaboration), Nucl.\ Phys.\ A \textbf{612}, 493 (1997)

\bibitem{shik}
S.~Shikazumi, T.~Maruyama, S.~Chiba, K.~Niita and A.~Iwamoto, Phys.\ Rev.\ C \textbf{63}, 024602 (2001)

\bibitem{frank}
J.D.~Frankland {\em et.~al.} (INDRA collaboration), Nucl.\ Phys.\ A \textbf{689}, 940 (2001)

\bibitem{bord}
B.~Borderie {\em et.~al.} (INDRA collaboration), Phys.\ Rev.\ Lett.\ \textbf{86}, 3252 (2001)

\end{thebibliography}
\end{document}